\documentclass[runningheads]{llncs}

\usepackage{graphicx}
\usepackage{bookmark}\usepackage{listings}
\usepackage{lmodern}
\usepackage{enumitem}
\usepackage{mathtools}
\usepackage[utf8]{inputenc}
\usepackage[T1]{fontenc}
\usepackage{lmodern}
\usepackage{amsmath, amssymb, amsthm}
\usepackage{url}
\usepackage{soul}
\usepackage{enumitem}
\usepackage{bm}
\usepackage{xcolor}
\hypersetup{
    colorlinks,
    linkcolor={red!50!black},
    citecolor={blue!50!black},
    urlcolor={blue!80!black}
}
\definecolor{linkcolor}{rgb}{0.65,0,0}
\definecolor{citecolor}{rgb}{0,0.65,0}
\definecolor{urlcolor}{rgb}{0,0,0.65}
\usepackage[nameinlink,noabbrev]{cleveref}

\setitemize{topsep=.9ex, itemsep=.1ex}
\usepackage{tikz}
\usepackage{cite}
\usepackage{caption}
\usepackage{booktabs}
\usepackage{relsize}
\usepackage{ifthen}
\usepackage{listings}
\usepackage{tikz-cd}
\usepackage{siunitx}
\usepackage{xspace}
\usepackage{todonotes}

\usepackage{xfp}  \usepackage{multirow}
\usepackage{tablefootnote}

\usepackage{algorithm,algpseudocode,mathdots}
\algdef{SE}[DOWHILE]{Do}{doWhile}{\algorithmicdo}[1]{\algorithmicwhile\ #1}

\usepackage{booktabs, pifont}

\newcommand\velusqroot{\ensuremath{\surd\text{élu}}}

\newcommand\Z{\mathbb Z}
\newcommand\F{\mathbb{F}}
\newcommand\Fp{\F_p}

\newcommand\fraka{\mathfrak{a}}
\newcommand\frakb{\mathfrak{b}}

\newcommand{\EC}{\ensuremath{\mathcal{E}}}

\newcommand{\xADD}{\ensuremath{\mathtt{xADD}}\xspace}
\newcommand{\xDBL}{\ensuremath{\mathtt{xDBL}}\xspace}
\newcommand{\ADDF}{\ensuremath{\mathtt{DIFF\_ADD}}\xspace}
\newcommand{\ctidh}{\textsf{CTIDH}\xspace}
\newcommand{\dcsidh}{\textsf{dCSIDH}\xspace}
\newcommand{\dctidh}{\textsf{dCTIDH}\xspace}
\newcommand{\tilO}{\widetilde{O}}
\newcommand{\dacshund}{\texttt{DACsHUND}\xspace}

\newcommand{\dachshund}{\ensuremath{\mathsf{DACsHUND}}\xspace}
\newcommand{\dachshunds}{\ensuremath{\mathsf{DACsHUND}}s\xspace}
\newcommand{\wombat}{\ensuremath{\mathsf{WOMBat}}\xspace}

\newcommand{\matryoshka}[2]{\ensuremath{
		\mathtt{Matryoshka}_{[{#1},{#2}]}}\xspace}
    
\renewcommand{\matryoshka}[2]{\if\relax\detokenize{#1}\relax
    \if\relax\detokenize{#2}\relax
      \ensuremath{\mathtt{Matryoshka}}\xspace
    \else
      \ensuremath{\mathtt{Matryoshka}_{[{#1},{#2}]}}\xspace
    \fi
  \else
    \ensuremath{\mathtt{Matryoshka}_{[{#1},{#2}]}}\xspace
  \fi
}

\newcommand{\M}{\ensuremath{\mathbf{M}}\xspace}     \newcommand{\Sq}{\ensuremath{\mathbf{S}}\xspace}     \newcommand{\A}{\ensuremath{\mathbf{a}}\xspace}

\crefname{ALG@line}{line}{lines}
\Crefname{ALG@line}{Line}{Lines}
\crefname{ALC@line}{line}{lines}
\Crefname{ALC@line}{Line}{Lines}
 \newif\ifsubmission
\submissionfalse

\begin{document}
\title{Hardened CTIDH: Dummy-Free and Deterministic CTIDH}
\ifsubmission
    \author{}
    \authorrunning{}
    \institute{}
\else
    \author{
    Gustavo~Banegas\inst{1}
    \and
    Andreas Hellenbrand\inst{2}
    \and
    Matheus Saldanha\inst{3}
    }
    \authorrunning{}
    \institute{
        Inria and Laboratoire d'Informatique de l'Ecole polytechnique,\\
        Institut Polytechnique de Paris, Palaiseau, France \\
        \email{gustavo@cryptme.in}
        \and
        RheinMain University of Applied Sciences Wiesbaden, Germany \\
        \email{andreas.hellenbrand@hs-rm.de}
        \and
        Universidade Federal de Santa Catarina, Florianópolis, Brazil \\
        \email{matheus.saldanha@posgrad.ufsc.br}
    }
\fi
\maketitle              \ifsubmission

\else
\vspace{-1ex}
  \makeatletter \begingroup \makeatletter \def\@thefnmark{$*$}\relax \@footnotetext{\relax Author list in alphabetical order; see \url{https://ams.org/profession/leaders/CultureStatement04.pdf}. 
    This work has been supported by the Federal Ministry of Research, Technology and Space (BMFTR) under the project 
    QUDIS (ID 16KIS2089), and by the HYPERFORM consortium, funded by France through Bpifrance, and by the France 
    2030 program under grant agreement ANR-22-PETQ-0008 PQ-TLS.

\def\ymdtoday{\leavevmode\hbox{\the\year-\twodigits\month-\twodigits\day}}\def\twodigits#1{\ifnum#1<10 0\fi\the#1}Date of this document: \ymdtoday.}\endgroup
\fi

\begin{abstract}
Isogeny-based cryptography has emerged as a promising post-quantum alternative, 
with CSIDH and its constant-time variants \ctidh and \dctidh offering efficient 
group-action protocols. However, \ctidh and~\dctidh rely on dummy 
operations in differential addition chains (DACs) and Matryoshka, which 
can be exploitable by fault-injection attacks. 
In this work, we present the first \emph{dummy-free} implementation of 
\dctidh. Our approach combines two recent ideas: \dacshund, which enforces 
equal-length DACs within each batch without padding, and a reformulated 
Matryoshka structure that removes dummy multiplications and validates all 
intermediate points. 
Our analysis shows that small primes such as $3,5,$ and $7$ severely restrict feasible 
\dacshund configurations, motivating new parameter sets that exclude them. 
We implement dummy-free \dctidh-2048-194 and \dctidh-2048-205, 
achieving group action costs of roughly $357{,}000$–$362{,}000$ $\Fp$-multiplications, 
with median evaluation times of $1.59$–$1.60$ (Gcyc). 
These results do not surpass \dctidh, but they outperform \ctidh by roughly $5\%$ 
while eliminating dummy operations entirely. Compared to dCSIDH, our construction 
is more than $4\times$ faster. 
To the best of our knowledge, this is the first \textit{efficient}
implementation of a CSIDH-like protocol that is simultaneously deterministic, 
constant-time, and fully dummy-free.  

\keywords{post-quantum cryptography \and isogeny-based cryptography \and CSIDH}
\end{abstract}
 
\section{Introduction}
\label{sec:intro}
In recent years, isogeny-based cryptography has attracted significant attention
from both mathematicians and cryptographers, due to its features such as 
non-interactive key exchange and compact key sizes. Following the cryptanalysis 
of SIKE~\cite{CastryckD23,MainoMPPW23,Robert23}, research on isogeny-based key 
exchange has shifted toward \textsf{CSIDH}~\cite{Castryck0MPR18}, which currently remains 
\emph{unbroken}. The attacks that compromised SIKE--based on the supersingular isogeny 
framework--do not apply to \textsf{CSIDH} or its variants, thereby preserving their
 relevance as viable post-quantum key exchange candidates.

Despite its resilience, \textsf{CSIDH} is relatively slow compared to other 
post-quantum schemes. Furthermore, achieving secure 
implementations requires countermeasures against side-channel 
attacks, which further increase computational overhead. 
To address these limitations, variants such as 
\ctidh~\cite{ctidh2021} and \dctidh~\cite{dctidh2025} have been 
proposed. These schemes improve performance by introducing more 
structured sets of isogeny paths and leveraging fixed parameter 
sets that simplify implementations. Both employ \emph{isogeny batching}
techniques, which simultaneously enhance security and 
performance. \ctidh achieves faster key exchange by introducing 
a new key space based on batches of isogenies, together with a 
constant-time algorithm for the \textsf{CSIDH} group action that synergizes 
with the new structure. Building on \ctidh, \dctidh adopts a more 
deterministic approach, refining the batching technique through the 
introduction of \emph{Widely Overlapping Meta-Batches} (WOMBats). 

Beyond performance, dummy operations introduce an attack surface 
for \emph{active} side-channels (fault injection): by targeting 
these redundant steps, an adversary can induce faults 
that desynchronize the control flow and leak secrets. This risk is not merely theoretical:
Campos, Kannwischer, Meyer, Onuki, and Stöttinger~\cite{CamposKMOS20} demonstrated 
fault attacks against dummy-padded isogeny computations, and 
more recently~\cite{ChiuL024} exploited 
dummies in implementations of \textsf{CSIDH} which are
constant-time. While batching in \ctidh raises 
the bar, their results indicate that \textbf{practical fault 
attacks remain feasible}, which motivates pursuing dummy-free 
techniques.

\paragraph{Contributions.}
In this work, we investigate in depth the use of \dacshund and 
dummy-free Matryoshka isogenies, and their combined role in enabling 
a fully dummy-free \dctidh implementation. 
Our primary goal is to produce an optimized variant of \ctidh/\dctidh 
that eliminates dummy operations while maintaining strong security properties.

\begin{enumerate}
    \item We analyze the \dacshund method for dummy-free DAC computations.  
    For each prime, we enumerate all possible DAC configurations and adapt 
    the \dctidh greedy parameter search to enforce equal-length \dachshunds 
    within each batch, thus avoiding dummy operations.  
    We evaluate the resulting configurations under different \dctidh settings 
    and quantify the performance impact.
    
    \item We implement the dummy-free Matryoshka isogeny approach and integrate 
    its cost into the greedy search process, enabling parameter optimization 
    that accounts for its specific constraints.
    
    \item We propose new \dctidh parameter sets that leverage these dummy-free 
    techniques for improved performance.  
    We focus on configurations that exclude the primes $3, 5, 7$, as they 
    are less compatible with \dacshund.

\end{enumerate}

\paragraph{Availability of software.}
Our implementation and greedy search scripts are available at
\begin{center}
	\url{https://github.com/AndHell/hardenedCTIDH}.
\end{center}

\paragraph{Related Work.}
Several works have sought to make \textsf{CSIDH} constant-time or deterministic.
For instance, in~\cite{michael_2019}, the authors address challenges such as point sampling and 
introduce the SIMBA technique. However, their approach still relies on dummy operations to compute 
isogenies. In parallel, other research has explored dummy-free constant-time methods, including two-point 
ladders and strategy-based scheduling of small-prime isogenies~\cite{ONUKI}. While these methods help 
mitigate timing leakage, they do not fully resolve batch-level DAC harmonization or eliminate the 
dummy padding inherent in Matryoshka. In \ctidh~\cite{ctidh2021}, the authors apply batching of 
isogenies using atomic blocks and Matryoshka to achieve a faster constant-time implementation of 
\textsf{CSIDH}. However, this approach is neither deterministic nor dummy-free.

Campos, Hellenbrand, Meyer, and Reijnders introduced \dctidh~\cite{dctidh2025}, 
a deterministic variant of \ctidh. Their central innovation is the use of 
\emph{WOMBats}, which combine overlapping batches with multiple isogenies per batch to enable 
efficient deterministic evaluation. Their implementation is highly optimized, both in terms of 
the number of finite-field operations per prime and the efficiency of those operations. 
Nevertheless, as the authors emphasize, \dctidh still relies on dummy operations in both 
Matryoshka isogenies and DAC padding. As a result, it is not dummy-free, leaving open the 
challenge of combining determinism, constant-time execution, and full dummy-freeness in a 
single construction. To address this, \dctidh proposed two potential directions: 
\emph{\dacshund} and \emph{dummy-free Matryoshka isogenies}. In this work, we investigate 
these approaches in detail, with the goal of achieving the first fully dummy-free variant of 
\dctidh.

Recent work has explored radical 3-isogenies as a replacement for small-degree 
isogenies in \textsf{CSIDH}-like protocols, reporting up to a $4\times$ speedup for 
dCTIDH~\cite{ChiDominguezOR25}. However, initiating a 3-isogeny chain still 
requires repeated sampling, which introduces probabilistic behavior. As a 
result, radical 3-isogeny chains cannot be made dummy-free and remain too 
costly in practice compared to 2-isogeny walks.

Addressing a related challenge in isogeny computation, 
Bernstein, Cottaar, and Lange~\cite{BernsteinCL24} revisit the problem of 
constructing differential addition chains, introducing new algorithms that 
minimize both chain length and computational overhead. 
Their work focuses on faster methods for finding minimum-length 
continued-fraction differential addition chains, significantly improving 
over previous search strategies. 
In our setting, we also rely on efficient DACs and employ a greedy search 
procedure to identify chains that satisfy the structural constraints imposed by 
\textsf{DACsHUND}. While their algorithms target global optimality in chain length, 
our approach prioritizes compatibility with batching and constant-time 
requirements, aiming at practical dummy-free instantiations within the 
\dctidh framework.

\section{Background}
\label{sec:background}
\subsection{Elliptic Curves and Isogenies}

Given a finite field $\Fp$, an \emph{elliptic curve} $E$ over $\Fp$ is defined by the equation
\[
    y^2 = x^3 + ax + b,
\]
where $a,b \in \F_p$ and $4a^3 + 27b^2 \neq 0$ to ensure the curve is nonsingular.  
Elliptic curves can also be expressed in alternative forms. 
For instance, \emph{Montgomery curves} constitute a special class of elliptic curves defined over $\Fp$ by
\[
    By^2 = x^3 + Ax^2 + x,
\]
where $A,B \in \Fp$ and $B(A^2 - 4) \neq 0$ guarantees nonsingularity. 
The group law on $E(\Fp)$ uses the point at infinity $\mathcal{O}$ as the identity element.
For a detailed treatment of elliptic curve theory and arithmetic, see~\cite{silverman2009arithmetic}.

\paragraph{Projective Coordinates.}
In practice, elliptic-curve arithmetic is often performed in 
\emph{projective coordinates} to avoid costly field inversions. 
An affine point $(x,y)$ is represented as $(X:Y:Z)$, corresponding to 
$(X/Z,\,Y/Z)$ when $Z \neq 0$, while the point at infinity $\mathcal{O}$ 
is given by $(0:1:0)$. 
This representation replaces inversions with a few additional multiplications, 
making additions, doublings, and isogeny evaluations both more efficient 
and easier to implement in constant time.

\paragraph{Isogenies.}
An \emph{isogeny} between elliptic curves, $\phi : E \to E'$, 
is a non-constant algebraic map that preserves the group law.  
Every isogeny is uniquely determined by its kernel, which is a finite subgroup of $E$.  
When working with Montgomery curves, these maps can be efficiently evaluated 
using only the $x$-coordinates of points, yielding major computational 
advantages for large-degree isogeny walks as required in 
\textsf{CSIDH}~\cite{Castryck0MPR18} and related protocols.  

The use of $x$-only arithmetic not only simplifies the application of 
Vélu’s formulas---the classical tool for computing an isogeny from its 
kernel---but also enables constant-time implementations via techniques such as 
the Montgomery ladder.  
For small odd prime degrees $\ell$, the kernel is usually generated 
by an $\Fp$-rational point of order $\ell$, which allows efficient construction 
of the corresponding quotient curve.  

The two main algorithms for evaluating such isogenies are 
\emph{Vélu's formulas}~\cite{velu1971isogenies} and 
\emph{$\velusqroot$} method~\cite{velusqr}.  
Both reduce the problem to computing a polynomial of the form
\begin{equation}
    h_S(X) = \prod_{s \in S} \big( X - x([s]P) \big),
\end{equation}
where $P$ is a point of order $\ell$, and $S$ is an index set determined by $\ell$.  
The main computational tasks are to determine the new Montgomery coefficient $A'$ of 
the image curve $E'$ and to evaluate the images of selected points under $\phi$.

In the classical Vélu approach, the index set is $S=\{1,2,\dots,(\ell-1)/2\}$; one computes 
$x([s]P)$ for all $s\in S$ and forms the product $h_S(X)$. This yields essentially linear
cost in $\ell$: about $4\ell$ $\Fp$-multiplications to update $A'$ and $2\ell$ per evaluated 
image point, i.e., overall $\tilO(\ell)$; Vélu is conceptually simple and practically 
optimal for small prime degrees.

For larger $\ell$, the $\velusqroot$\footnote{Pronounced ``square-root Vélu''.} algorithm applies a baby-step/giant-step decomposition on 
the odd-index set $S=\{1,3,5,\dots,\ell-2\}$ via $S\leftrightarrow (U\times V)\cup W$, 
obtaining $h_S$ as $h_W$ times a resultant involving $h_U$ and a polynomial derived from 
$h_V$. This reorganizes the arithmetic to $\tilO(\sqrt{\ell})$. Vélu is a special 
case of $\velusqroot$ with $U=V=\varnothing$.

\begin{remark}
We note that both \ctidh and~\dctidh rely on
\emph{Matryoshka} ``isogenies,'' a technique to enforce uniform
evaluation costs across batches of primes. Because this construction is
central to \ctidh,~\dctidh, and also to our work, we provide a more detailed
discussion in \S\ref{sec:back:techniques}.
\end{remark}

\subsection{\textsf{CSIDH}}

Introduced by~\cite{Castryck0MPR18} in 2018, \textsf{CSIDH} is a 
non-interactive key exchange protocol based on the action of the 
ideal class group of an imaginary quadratic order on a set of 
supersingular elliptic curves defined over a prime field $\Fp$.  
This class group action is realized through chains of isogenies 
between elliptic curves, each of small odd prime degree 
$\ell_i$ dividing $p + 1$.

The protocol operates on a restricted set $\EC$ of 
supersingular elliptic curves $E / \Fp$ whose endomorphism ring is 
isomorphic to $\Z[\sqrt{-p}]$, with all curves having exactly 
$p + 1$ points.  
For a prime $p$ of the form
\[
    p + 1 = 2^f \cdot g \cdot \prod_{i=1}^n \ell_i,
\]
where $f \geq 2$, $g$ a small cofactor, and the $\ell_i$ are small, 
distinct odd primes, 
the group structure of $\EC(\Fp)$ admits a torsion decomposition 
enabling efficient computation of $\ell_i$-degree isogenies.

The underlying group action is defined as follows: a secret key is a
 vector $(e_1, \dots, e_n)$ with $e_i \in [-m_i, m_i]$, representing 
 the ideal class
\[
    \fraka = \prod_{i=1}^n \ell_i^{e_i}.
\]
Its action on a fixed base curve $E_0$ is computed as a walk in the 
$\ell_i$-isogeny graph, with each step corresponding to an 
$\ell_i$-isogeny in the forward or backward direction according to 
the sign of $e_i$.  
The resulting curve $E' = \fraka * E_0$ is the public key.

\textsf{CSIDH} is \emph{commutative}: for secret keys $\fraka$ and $\frakb$,
\[
    \fraka * (\frakb * E_0) = \frakb * (\fraka * E_0),
\]
allowing both parties to derive the same shared secret curve 
without interaction.

Security relies on the \emph{isogeny path-finding problem}: given two 
supersingular curves $E$ and $E'$ over $\Fp$ with the
same $\Fp$-rational endomorphism ring $\mathcal{O}$, find an explicit 
$\Fp$-rational isogeny $\phi : E \to E'$ of smooth degree.  
This problem is believed to be hard for both classical and 
quantum algorithms when instantiated with sufficiently large 
$p$ and appropriate parameters.  
Quantum security analysis remains active, with recent work 
suggesting that primes of at least 2048 bits may be required 
for conservative approaches.

\subsection{Constant-Time Isogeny Diffie–Hellman (\ctidh)}  
The \ctidh~\cite{ctidh2021} variant removes timing side
channels by ensuring that all isogeny walks execute in constant time.
Instead of conditionally applying an $\ell_i$-isogeny based on the exponent $e_i$,
\ctidh introduces a batching strategy with a redefined key space. A batch
is defined as $\mathcal{B}_i = \{\ell_{i,1}, \dots, \ell_{i,N_i}\}$, where all primes in
$\mathcal{B}_i$ are handled collectively. An $\ell_{i,j}$-isogeny is then computed
as an $\ell_{i,N_i}$-isogeny through \emph{Matryoshka isogenies}, which conceal
the degree of each isogeny by padding smaller ones with dummy computations
so that every evaluation matches the cost of an $\ell_{i,N_i}$-isogeny for each batch $B_i$.

To further mitigate leakage, \textsf{CTIDH} assigns a bound $m_i$ to each batch, 
prescribing a fixed number of isogeny evaluations. When the required number of 
evaluations is smaller than $m_i$, dummy isogenies are inserted so that every 
batch always performs exactly $m_i$ evaluations. This masks timing variations 
across batches. However, it does not protect against fault attacks, since the 
dummy operations themselves remain a potential target.

The dominant cost of an isogeny evaluation lies in computing its kernel polynomial,
which involves scalar multiplications by different prime factors and would otherwise
lead to timing variations. To mitigate this, \ctidh employs 
\emph{differential addition chains} (DACs). By padding shorter chains with dummy
 operations, all scalar multiplications are forced to cost the same, analogous to the 
Matryoshka approach used for isogeny evaluations.

Finally, performance improvements arise from assigning bounds $m_i$ to entire batches
rather than to individual primes. This yields a larger combinatorial key space:
\[
\#K_{N,M} = \prod_{i=1}^B \Phi(N_i, m_i),
\quad \Phi(x,y) = \sum_{k=0}^{\min\{x,y\}} \binom{x}{k} \binom{y}{k} 2^k,
\]
where $\Phi(x,y)$ counts integer vectors in $\mathbb{Z}^x$ with $\ell_1$-norm at most $y$.

\subsection{\dcsidh: Deterministic and Dummy-Free \textsf{CSIDH}}

The \dcsidh, or \texttt{secsidh}, variant~\cite{secsidh} was introduced 
as a high-security implementation of \textsf{CSIDH} that simultaneously achieves 
\emph{determinism} and \emph{dummy-freeness}. Unlike \ctidh, which 
relies on dummy operations to enforce constant-time behavior, 
\dcsidh eliminates both randomness and dummy padding by restricting 
the key space to exponents $e_i \in \{-1, 1\}$. This restriction 
ensures that every isogeny degree is used exactly once in a fixed 
direction, providing determinism in both point sampling and isogeny 
evaluation.

From a security perspective, determinism provides a stronger defense 
against fault attacks. However, from a performance standpoint, this comes 
at a significant cost: eliminating dummies removes batching flexibility, 
and determinism requires larger parameter sizes (typically starting at 
2048-bit primes) to maintain security. As a result, benchmarks show that 
\dcsidh runs approximately $3$ to $5$ times slower than probabilistic 
\ctidh at equivalent parameter sizes.

\subsection{\dctidh: Deterministic \ctidh}

While \textsf{CSIDH} offers an elegant algebraic structure and promising 
post-quantum security, its reference design is vulnerable to practical implementation
issues most notably timing and fault attacks. To address these challenges, 
\dctidh~\cite{dctidh2025} was introduced as a refinement of \ctidh, 
enhancing the original protocol with deterministic evaluation, stronger 
side-channel resistance, and improved performance.

The \dctidh scheme is a deterministic variant of
\ctidh that resolves the reliance on probabilistic point
sampling and non-deterministic isogeny evaluation. Its key innovation is
the introduction of \emph{Widely Overlapping Meta-Batches} (WOMBats),
which combine two complementary batching ideas: \emph{multiple isogenies
per batch} and \emph{overlapping batches}.

In the original \ctidh, exactly one isogeny is computed per
batch in order to avoid secret-dependent behavior. This constraint
limits efficiency, since even if the secret key requires several
isogenies from the same batch, only one can be evaluated. In contrast,
if we restrict secret exponents to unitary values $e_i \in \{-1,1\}$,
then multiple isogenies of distinct degrees can be computed safely
within a single batch. For a batch
$\mathcal{B}_i = \{\ell_{i,1}, \dots, \ell_{i,N_i}\}$, we can choose any number
$M_i \leq N_i$ of distinct degrees, evaluating $M_i$ isogenies via
$M_i$ calls to $\mathrm{Matryoshka}[\ell_{i,1},\ell_{i,N_i}]$. 

This
significantly reduces the number of total isogenies needed, as the key
space grows combinatorially:
\[
   \Psi(N_i,M_i) = \binom{N_i}{M_i}\cdot 2^{M_i} \quad \text{or} \quad 
    \Psi_{\mathrm{dummy}}(N_i,M_i) = \sum_{j=0}^{M_i} \binom{N_i}{j}\cdot 2^j
\]
if dummy isogenies are allowed (i.e. $e_i \in \{-1,0,1\}$).

Another approach to enlarge the key space and improve efficiency is to
use batches that overlap in some of their prime factors. Suppose the
first batch is
$\mathcal{B}_1 = \{\ell_1,\dots,\ell_{N_1}\}$. Instead of defining
$\mathcal{B}_2 = \{\ell_{N_1+1},\dots,\ell_{N_1+N_2}\}$, we let the batches share
$\omega_{1,2}$ primes:
\[
   \mathcal{B}_2 = \{\ell_{N_1-\omega_{1,2}+1}, \dots, \ell_{N_1+N_2-\omega_{1,2}}\}.
\]
This overlapping structure amplifies the combinatorial growth of the
key space without requiring a proportional increase in the number of
isogeny evaluations. To ensure determinism, the bounds $M_1,M_2$ must
satisfy $M_1+M_2 \leq N_1+N_2-\omega_{1,2}$, preventing multiple
isogenies from being applied to the same degree.

\textsf{dCTIDH} combines the two techniques above into WOMBats. A WOMBat
$\mathcal{W}=\{\ell_{i,1},\dots,\ell_{i,N}\}$ with bound $M$ is evaluated as $M$
overlapping batches
\[
   \mathcal{B}_1 = \{\ell_1,\dots,\ell_{N-M+1}\}, \quad
   \mathcal{B}_2 = \{\ell_2,\dots,\ell_{N-M+2}\}, \quad \dots \quad
   \mathcal{B}_M = \{\ell_M,\dots,\ell_N\}.
\]
Each $\mathcal{B}_j$ overlaps in $N-M$ primes with its neighbors, and exactly one
isogeny is computed from each, realized as a Matryoshka isogeny
$\mathrm{Matryoshka}[\ell_j,\ell_{N-M+j}]$. In this way, the WOMBat
structure deterministically covers all possible distributions of $M$
distinct isogeny degrees, while guaranteeing constant computational
cost. The resulting key space of $N_\mathcal{W}$ disjoint WOMBats is
\[
   \prod_{i=1}^{N_\mathcal{W}} \Psi(N_i,M_i)
   = \prod_{i=1}^{N_\mathcal{W}} \binom{N_i}{M_i}\cdot 2^{M_i}.
\]

To mitigate timing leakage, \dctidh employs \emph{DACs},
which pad shorter chains with dummy steps within each WOMBat
to achieve constant-time scalar multiplication as previous mentioned. 
Consequently, even though \dctidh eliminates \emph{randomness} during evaluation, 
\emph{the DAC and Matryoshka computations still incorporate dummy steps to
maintain constant-time execution.}

\subsection{Techniques in \textsf{CSIDH}-like Schemes}\label{sec:back:techniques}

Efficient implementations of \textsf{CSIDH} and its variants rely on specialized 
techniques that simultaneously ensure constant-time execution and improve 
the performance of scalar multiplications and isogeny evaluations.  
Among the most important are \emph{Differential Addition Chains (DACs)}, 
which realize scalar multiplications in constant time using only 
$x$-coordinates, and \emph{Matryoshka isogenies}, which enable 
constant-time evaluation of isogenies while reducing computational cost 
through the exploitation of nested structures within isogeny chains. 

\paragraph{Differential Addition Chains (DACs).}
Differential Addition Chains (DACs) are algorithmic frameworks for 
scalar multiplication on elliptic curves, particularly in the Montgomery model, 
where only $x$-coordinates are used.  
By avoiding full group operations and secret-dependent branching, DACs enable 
constant-time and side-channel-resistant implementations—an essential feature 
in isogeny-based cryptography where points are ephemeral and curves evolve 
along isogeny walks.

\begin{definition}[Differential Addition Chain]
A \emph{differential addition chain} for an integer $n$ is a sequence 
\( 1 = c_0,\, c_1,\, \dots,\, c_r = n \)
such that for each $i \in \{1, \dots, r\}$ there exist indices $j,k < i$ with
\[ c_i = c_j + c_k, \text{and } c_j - c_k \in \{0, c_0, c_1, \dots, c_{i-1}\}. \]
\end{definition}

In other words, each new sum in the chain must correspond to a difference 
already present in the chain (or zero).  

\begin{example}
A differential addition chain for $29$ is
\( 1,\, 2,\, 3,\, 5,\, 8,\, 13,\, 21,\, 29, \)
since, for instance, $13 = 8 + 5$ with difference $8 - 5 = 3 \in \{1,2,3,5,8\}$.
\end{example}

In this work we focus on the subclass of \emph{continued-fraction DACs}, 
which admit a compact bitstring encoding. 
For simplicity, we use the terms \emph{DAC} and \emph{continued-fraction DAC} 
interchangeably throughout.

\begin{definition}[Continued-fraction DAC]
Let $(a_2,b_2,c_2), \dots, (a_r,b_r,c_r)$ be a sequence of triples with 
$n \geq 3$, $(a_2,b_2,c_2) = (1,2,3)$, $c_r = n$, and for each $i \geq 3$:
\[
    (a_i,b_i,c_i) =
    \begin{cases}
        (b_{i-1}, c_{i-1}, c_{i-1}+b_{i-1}), & \text{if $f_i = 0$}, \\[4pt]
        (a_{i-1}, c_{i-1}, c_{i-1}+a_{i-1}), & \text{if $f_i = 1$},
    \end{cases}
\]
with $c_i = a_i + b_i$.  
Then the continued-fraction DAC is the sequence
\( 1,\, 2,\, c_2,\, \dots,\, c_r = n. \)
\end{definition}

\begin{example}
A continued-fraction DAC for $13$ admits the compressed bitstring 
\( f = 11110. \)
\end{example}

In the Montgomery model, scalar multiplication $[k]P$ can be realized by iterating 
only differential operations:
\[
    \ADDF(P, Q, P-Q)
    \quad\text{and}\quad
    \xDBL(P),
\]
while tracking the differential $P-Q$.  
This makes the procedure fully deterministic and constant-time.
Algorithm~\ref{alg:dac} illustrates how scalar multiplication by $n$ 
can be carried out using a compressed DAC bitstring, relying solely 
on the two fundamental operations $\xDBL$ and $\ADDF$.

\begin{algorithm}
    \caption{\texttt{DAC} — Scalar multiplication via a compressed DAC}
    \label{alg:dac}
    \begin{algorithmic}[1]
    \Require Point $P$, compressed DAC $f_3,\dots,f_r$ for $n$
    \Ensure $[n]P$
    
    \State $X_0 \gets P$
    \State $X_1 \gets \xDBL(P)$
    \State $X_2 \gets \ADDF(P, P, X_1)$
    
    \For{$i = 3$ to $r$}
        \If{$f_i = 0$}
            \State $(X_0, X_1, X_2) \gets (X_1, X_2, \ADDF(X_0, X_1, X_2))$
        \Else
            \State $(X_0, X_1, X_2) \gets (X_0, X_2, \ADDF(X_1, X_0, X_2))$
        \EndIf
    \EndFor
    
    \State \Return $X_2$
    \end{algorithmic}
\end{algorithm}
Note that two DACs corresponding to different integers incur the same 
computational cost whenever their compressed representations have
the same length, regardless of the integers themselves.

\begin{remark}
Compared to the classic Montgomery ladder 
(which is also constant-time), continued-fraction DACs compress 
structured additions/doublings for fixed small $\ell$ and 
integrate more naturally with batch scheduling, which is why \ctidh/\dctidh prefer DACs for 
kernel generation.
\end{remark}

In \textsf{CSIDH}-like protocols, DACs are used to compute kernel 
generators for isogenies. Secret keys are exponent vectors $(e_1,\dots,e_n)
$ indicating how many times an isogeny of a specific degree is applied. 
Each scalar multiplication $[\ell_i]P_i$ (for small primes $\ell_i$)
is performed using a fixed DAC, ensuring constant-time execution.

In \ctidh and \dctidh, DACs are precomputed according
to the allowed exponent bounds, and scalar multiplications are often 
\emph{batched} to reuse intermediate results. However, the length of these
DACs—and therefore the computational cost of a multiplication by 
$\ell_i$—depends directly on $\ell_i$. To keep the isogeny degree 
$\ell_{i}$ secret, \textsf{CTIDH} enforces constant-time multiplications
for all factors within a batch $B$. This is done by precomputing an 
optimal DAC for each $\ell_{i} \in B$ and padding it with dummy 
steps if necessary, so that multiplication by any cofactor from 
$B$ requires the same number of operations as the largest $\ell_i$
in the batch. 

However, since dummy padding is normally
applied to maintain constant-time execution, it can leave room
for active attacks, such as fault injection. The \textsf{dCTIDH} 
scheme addresses this issue with \emph{\dacshund}, a 
technique for dummy-free DAC 
evaluation, which we explore later in this work.

\paragraph{Matryoshka isogenies.}
As previously mentioned, the computational cost of evaluating isogenies via 
Vélu’s formulas or $\velusqroot$ grows respectively as 
$\tilO(\ell)$ and $\tilO(\sqrt{\ell})$ in the isogeny degree $\ell$.  
Since in \textsf{CSIDH}-like protocols one must evaluate isogenies of different prime degrees, 
these costs naturally vary across primes, 
potentially leaking information and complicating optimization.
Additionally, when primes are grouped in batches, such as in \ctidh and \dctidh,
isogeny evaluations must also cost the same within each batch.
To address this, \ctidh introduces the notion of 
\emph{Matryoshka isogenies}, 
a technique that enforces uniform evaluation cost across a batch of primes.

The core idea is to impose a ``nested’’ evaluation structure on the kernel 
polynomial
\begin{equation}
    h_S(X) = \prod_{s \in S} \big( X - x([s]P) \big),
\end{equation}
where $S = \{1, 2, \dots, (\ell-1)/2\}$ in the Vélu case, 
or $S = \{1, 3, 5, \dots, \ell-2\}$ in the $\surd\text{élu}$ case.
For Vélu’s method, this amounts to cycling through 
the multiples $[s]P$, generating and evaluating $h_S(X)$ on the fly.  
Once the loop reaches $(\ell-1)/2$, one can continue appending 
\emph{dummy iterations}, thereby aligning the total number of operations 
to that required by the largest prime $\ell_{\max}$ in the batch.  
In this way, any $\ell$-isogeny in the batch can be evaluated 
at the uniform cost $\tilO(\ell_{\max})$.

The same concept extends to $\velusqroot$ evaluations.
In this case, the index set $S \;\longleftrightarrow\; (U \times V) \cup W$ is split into a box
$U \times V$ and a leftover set $W$. Then, $h_S(X)$ can be computed 
by multiplying $h_W(X)$ with the resultant of $h_U(X)$ and a 
polynomial derived from $V$, with all sets $U$, $V$,
and $W$ having size $\tilO(\sqrt{\ell})$.

To apply a Matryoshka structure, $U$ and $V$ are chosen according 
to the smallest degree in the batch, while $W$ is padded 
according to the largest.  
This ensures a uniform evaluation cost across primes in the batch, 
albeit with some efficiency loss since the parameters $U, V, W$ 
are no longer optimally tuned for each $\ell$.

We denote by $\text{Matryoshka}[\ell_i, \ell_j]$ a computation that performs any 
isogeny of degree $\ell \in [\ell_i, \ell_j]$ at the cost of $\ell_j$, 
whether using Vélu or $\surd\text{\'elu}$ as appropriate.  
This nested framework makes it possible to batch isogeny evaluations 
without leaking degree information, while still achieving sublinear 
performance when $\velusqroot$ is applicable.  
For further details, see~\cite{ctidh2021,velusqr}.

\section{DACsHUND}
\label{sec:dacshund}
In \textsf{dCTIDH}, key generation requires computing a sequence of scalar 
multiplications \([\ell_i]P\) for a fixed set of primes 
\(\ell_1, \dots, \ell_n\). These multiplications are carried out on Montgomery curves using 
$x$-only arithmetic (\xADD{}, \xDBL{}) and differential addition chains (DACs).  
Implementations must be side-channel resistant, deterministic, and ideally batched 
to maximize performance.  

In constant-time settings, the minimal DAC for each prime generally has a different length. 
To equalize the execution flow, previous approaches required padding shorter DACs with 
dummy operations so that all scalar multiplications within a batch completed in the same number of steps.  
While effective, this introduces redundancy and increases susceptibility to certain advanced 
fault-injection attacks. To overcome this limitation, we introduce 
\emph{\dacshund} (Differential Addition Chain Having Unnecessities Needed for Dummy-freeness), 
originally proposed in the future work of the \textsf{dCTIDH} paper, which enables 
dummy-free DAC execution.

\begin{definition}[\dacshund]
Let $\{\mathcal{B}_1,\dots,\mathcal{B}_n\}$ be a family of $n$ batches, where each batch $\mathcal{B}_i$
consists of $N_i$ primes: $\mathcal{B}_i = \{\ell_{1,i}, \dots, \ell_{N_i,i}\}$ with 
$\ell_{1,i} \leq \dots \leq \ell_{N_i,i}$.  
Each prime $\ell_{j,i}$ has an associated set $\mathcal{D}_{j,i}$ of admissible DAC lengths.  
The configuration $\{\mathcal{B}_1,\dots,\mathcal{B}_n\}$ is a valid \emph{\dacshund} 
if, for every batch $\mathcal{B}_i$, the intersection 
\(\bigcap_{j=1}^{N_i} \mathcal{D}_{j,i}\) is non-empty.
\end{definition}

Intuitively, the idea is to partition the primes into batches such that all DACs in a batch share 
at least one common length. This eliminates the need for dummy padding while preserving constant-time execution.  
Algorithm~\ref{alg:isvalid-dacshund} formalizes the batch validation procedure.  
This general framework not only supports \textsf{dCTIDH}, but can also be applied to 
related protocols such as \textsf{CTIDH}.  

\begin{example}
Consider a batch $\mathcal{B}_{1} = \{11, 13, 17, 19\}$.  
The corresponding DAC sets are:  
\[
\begin{aligned}
\mathcal{D}_{1,1} &= \{3,4,8\}, \\
\mathcal{D}_{2,1} &= \{3,4,5,10\}, \\
\mathcal{D}_{3,1} &= \{4,5,7,14\}, \\
\mathcal{D}_{4,1} &= \{4,5,6,8,16\}.
\end{aligned}
\]
Since their intersection is $\{4\}$, this is a valid \emph{\dacshund} configuration.  
However, if prime $5$ is added, its DAC set $\{1,2\}$ leads to an empty intersection, 
invalidating the batch.
\end{example}

\begin{algorithm}[H]
    \caption{\texttt{IsValidDACsHUND} --- Validation of DACsHUND Compatibility}
    \label{alg:isvalid-dacshund}
    \begin{algorithmic}[1]
    \Require Batch sizes \(N = (N_1, \dots, N_B)\), number of batches \(B\), prime list \(\mathcal{P}\)
    \Ensure \texttt{True} if valid; \texttt{False} otherwise
    
    \State Partition \(\mathcal{P}\) into batches \(\mathcal{P}^{(1)}, \dots, \mathcal{P}^{(B)}\) of sizes \(N_1, \dots, N_B\)
    
    \For{$i = 1$ to $B$}
        \State $I \gets \bigcap_{p \in \mathcal{P}^{(i)}} \texttt{DACsHUND}[p]$
        \If{$I = \emptyset$}
            \State \Return \texttt{False}
        \EndIf
    \EndFor
    
    \State \Return \texttt{True}
    \end{algorithmic}
\end{algorithm}

\paragraph{\dacshund Map.} The first step in building a \dacshund configuration is to enumerate all admissible DACs 
for each prime in the range of interest. Instead of storing only the shortest DAC, 
we record every possible DAC length and its corresponding representation. 
This yields a map $\texttt{DACsHUND}$ associating each prime $p$ with its set of DAC lengths.  
For example, $\texttt{DACsHUND}[13] = \{3,4,5,10\}$.

We adopt a straightforward \emph{brute-force} strategy: 
enumerating all possible compressed DAC representations 
up to a prescribed length (e.g., $16$), and testing each candidate 
to verify whether it corresponds to a valid prime.  
Although this approach does not exploit optimized DAC search 
methods~\cite{BernsteinCL24}, the search space remains sufficiently small that 
an exhaustive traversal can be completed in a small time frame.

\subsection{Searching Batch Configurations}

With $\texttt{DACsHUND}$ in place, the next step is to search for valid batch configurations.  
The \textsf{dCTIDH} batch search builds on the greedy
strategy of \textsf{CTIDH} and is defined by three parameters:  
the number of batches $B$, the batch size vector $N = (N_1,\dots,N_B)$
specifying the number of primes per batch,  
and the bound vector $M = (M_1,\dots,M_B)$ that ensures the
resulting configuration spans a sufficiently large key space.

\paragraph{Initialization.} The standard dCTIDH greedy initialization assigns equal size to all batches 
(\(N_i = n/B\) with $\sum N_i = n$), but this often produces invalid \dacshund 
configurations with empty intersections.  
To address this, we construct an initial configuration iteratively: starting with 
\(N = (1, \dots, 1)\), we cycle through the batches, incrementing one $N_i$ at a time, 
and accept the update only if the resulting configuration is \dacshund-valid.  
This continues until all primes are allocated.  
The procedure is shown in Algorithm~\ref{alg:find-initial-batch-sizes}.

\begin{algorithm}[H]
    \caption{\texttt{FindInitialBatchSizes} --- Search for Valid Initial Configurations}
    \label{alg:find-initial-batch-sizes}
    \begin{algorithmic}[1]
    \Require Number of batches \(B\), prime list \(\mathcal{P}\)
    \Ensure Batch size tuple \(N\) if valid; \texttt{None} otherwise
    
    \State Initialize \(N \gets (1, \dots, 1) \in \mathbb{Z}^B\)
    
    \While{\(\sum_{i=1}^{B} N_i < |\mathcal{P}|\)}
        \State \(\Delta \gets \texttt{False}\)
        \For{$i = 1$ to $B$}
            \State Let \(N' \gets N\) with \(N'_i \gets N_i + 1\)
            \If{ \Call{IsValidDACsHUND}{$N', B, \mathcal{P}$}}
                \State \(N \gets N'\), \(\Delta \gets \texttt{True}\)
            \EndIf
        \EndFor
        \If{\(\Delta = \texttt{False}\)}
            \State \Return \texttt{None}
        \EndIf
    \EndWhile
    
    \State \Return \(N\)
    \end{algorithmic}
\end{algorithm}

\paragraph{Greedy search.} The greedy algorithm modifies a configuration by 
decreasing the size $N_i$ of one batch $B_i$ and increasing the size of 
another $B_j \neq B_i$.  
This is repeated while exploring feasible bounds $M_i$ for each batch.  
To integrate \dacshund, we introduce a validation step at each modification to
ensure that the new batch configuration preserves non-empty DAC intersections.  
If multiple DAC lengths are available, the smallest one is selected to minimize 
scalar multiplication cost. The cost function is thus adapted to consider the 
shortest valid DAC from the intersection of each batch.  

\begin{remark}
Small primes such as $3$, $5$, and $7$ have very restricted DAC sizes 
(e.g., $\mathcal{D}_{3} = \{0\}$). Their inclusion can yield inefficient 
configurations under \dacshund constraints. For this reason, we also explore 
configurations excluding these primes and substituting them with larger ones 
to assess the performance trade-offs.
\end{remark}

\section{Dummy-Free Matryoshka}
\label{sec:matryoshka}

As outlined in \S\ref{sec:back:techniques}, both \ctidh and \dctidh
employ the Matryoshka structure to conceal the true degree of an isogeny 
within a batch. In this setting, an isogeny of degree $\ell_k$ contained in 
a batch $(\ell_l, \ell_r)$ is evaluated at the uniform cost of an 
$\ell_r$-isogeny. The classical construction proceeds as follows. One first computes the 
sequence of points
\[
    P, [2]P, \dots, \Bigl[\tfrac{\ell_r-1}{2}\Bigr]P,
\] 
and from these builds the kernel polynomial. The polynomial is factored into 
two parts: the \emph{real factors},
\[
    \prod_{i=0}^{(\ell_k-1)/2} \bigl(x - x([i]P)\bigr),
\]
which correspond to the actual $\ell_k$-isogeny, and the \emph{dummy factors},
\[
    \prod_{i=(\ell_k-1)/2+1}^{(\ell_r-1)/2} \bigl(x - x([i]P)\bigr),
\]
which pad the cost up to $\ell_r$ and thereby hide the true degree $\ell_k$.

This dummy-based approach introduces two distinct entry points for 
fault-injection attacks. First, the dummy multiplications in the kernel 
polynomial may be distinguishable from real ones, enabling targeted faults. 
Second, the unused multiples
\[
    \Bigl[\tfrac{\ell_k-1}{2}+1\Bigr]P, \dots, \Bigl[\tfrac{\ell_r-1}{2}\Bigr]P,
\]
although computed, are never required by the true kernel and thus create 
additional leakage channels. 

To address these vulnerabilities,~\cite{dctidh2025} introduced a modified 
Matryoshka structure. Their refinement eliminates dummy multiplications by 
reformulating the kernel product so that redundant terms cancel out 
algebraically, rather than being introduced explicitly. Furthermore, the 
unused multiples are validated against their expected relations, preventing 
an adversary from exploiting them as a source of leakage. This restructuring 
preserves the constant-time nature of Matryoshka while \emph{significantly reducing 
its exposure to fault attacks}.

\subsection{Matryoshka 2.0}

The idea described in \cite[Appendix A]{dctidh2025}, eliminates dummy operations entirely 
while retaining the same cost profile. 
The key observation is that for any point $P$, we have $x([i]P) = x([\ell - i]P)$.
This symmetry allows the algorithm to verify that every multiple's $x$-coordinate 
must be computed correctly, since each will appear twice.

\Cref{alg:matry2} shows the full computation of the kernel polynomial $h$ 
using the dummy-free Matryoshka approach, as described in \cite{dctidh2025}. 
Instead of inserting dummy factors, Matryoshka~2.0 replaces them with real 
multiplications of a modified form:
\[
    x = \tfrac{1}{2}x([i]P) - \alpha \cdot \tfrac{1}{2}x([i]P),
\] 
where $\alpha$ is chosen in constant time to be $-1$ if the value 
$x([i]P)$ has already appeared for some $j<i$, and $1$ otherwise. 
Thus, lines~\ref{alg:matry2:comp:start} to~\ref{alg:matry2:comp:end} 
carry the same information as checking whether $i > \tfrac{\ell_k - 1}{2}$ 
to determine if a dummy operation needs to be computed in the original version.

This achieves two crucial properties: uniformity of computation, 
since every iteration performs a real multiplication of the same 
cost, leaving no distinction between \textit{real} and \textit{dummy} steps; 
and the absence of unused data, since all multiples $x([i]P)$ are 
incorporated into the product, eliminating the risk of computing 
unnecessary points.

\begin{algorithm}[!h]
    \caption{Matryoshka 2.0 (based on \cite{dctidh2025})}\label{alg:matry2}
    \begin{algorithmic}[1]
    \Require A degree $\ell_k$, a batch $[\ell_l, \ldots \ell_r]$ 
            and a point $P$ such that $\ell_k \cdot P = \mathcal{O}$
    \Ensure The kernel polynomial $h(x)$ for $\phi: E \to E/\langle P \rangle$
    
    \item[]
    \State $b_k \leftarrow \frac{\ell_k - 1}{2}, b_l \leftarrow \frac{\ell_l - 1}{2}, b_r \leftarrow \frac{\ell_r - 1}{2}$
    \State $t \leftarrow b_r - b_l$
    \State Compute ($x$-coordinates of) $\{P,[2]P, \dots, [b_r]P \}$.
    \State $h(x) \leftarrow 1$
    
    \item[]
    \For{$i \in [1, \ldots, b_l ]$} \Comment{compute the \textit{linear} part up to $b_l$}
        \State $m \leftarrow x([i]P)$
        \State $h(x) \leftarrow h(x) \cdot (x - m)$
    \EndFor
    
    \item[]
    \For{$i \in [b_l + 1, \ldots, b_r ]$}
        \State $m \leftarrow \frac{1}{2}x([i]P)$
        \State $\alpha \leftarrow 1$
        \For {$j \in [(b_l + 1 - t), \ldots, (i - 1) \}$} \Comment{checks if $x[iP]$ has appeared already}\label{alg:matry2:comp:start}
            \State $\alpha \leftarrow \alpha \cdot$ \Call{cCompare}{$x([i]P)$, $x([j]P)$} \Comment{returns -1 if so}
        \EndFor	\label{alg:matry2:comp:end}
        \State $h_1(x) \leftarrow h(x)\cdot(x - m)$\label{alg:matry2:h1}
        \State $h_2(x) \leftarrow h(x)\cdot \alpha \cdot m$
        \State $h(x) \leftarrow h_1(x) - h_2(x)$\label{alg:matry2:h1h2}
    \EndFor 
    
    \item[]
    \State \Return $h(x) \leftarrow x^{b_k - b_r} \cdot h(x)$\label{alg:matry2:correction} \Comment{fix the degree}
    \end{algorithmic}
\end{algorithm}

The original Matryoshka implementation in \ctidh (and \dctidh) uses projective space 
to represent $x$-only points as $(X:Z)$, thereby avoiding costly inversions. 
As in Vélu’s formulas, the kernel polynomial must be evaluated at $\tfrac{h(1)}{h(-1)}$ 
to compute the codomain coefficient $A'$. In the projective setting, the evaluations at $1$ 
and $-1$ are directly integrated into the implementation.

To adapt \Cref{alg:matry2} to projective space, we replace the affine 
expression $m = \tfrac{1}{2}x([i]P)$ with its projective equivalent. 
Writing $[i]P = (X_i : Z_i)$, we obtain
\[
    \frac{m_x}{m_z} = \frac{X_i}{2 \cdot Z_i}.
\]
Accordingly, we updated in lines~\ref{alg:matry2:h1}--\ref{alg:matry2:h1h2} 
with the following 
\[
    \frac{h_x}{h_z} \;=\; 
    \frac{h_x \cdot \bigl((\alpha \cdot m_x) + m_x - m_z\bigr)}{h_z \cdot \bigl((\alpha \cdot m_x) + m_x + m_z\bigr)}.
\]
The \texttt{cCompare} routine must also be modified to compare projective points, 
increasing its cost to $2\M$. 
Finally, the degree correction step in line~\ref{alg:matry2:correction} simplifies 
to a constant-time sign flip of $h_z$.

Igonoring additions, the computation of one $\matryoshka{\ell_l}{\ell_r}$-isogeny is thereby increased
by $\sum_{i=1}^t (t-1+i) \cdot 2 \M$, with $t = ((\ell_r - 1) / 2) - ((\ell_l - 1)/2)$, 
compared to the dummy based version.

\subsection{Matryoshka 1.414 (\velusqroot) }

For the Matryoshka\footnote{We called Matryoshka 1.414 since $\sqrt{2} \approx 1.414$.} 
variant using \velusqroot, \emph{\Cref{alg:matry2} cannot be applied directly}, 
since not all multiples $[i]K$ required for comparison are available 
due to the index system that splits the computation into
$U \times V \cup W$. However, we can exploit the structure of Matryoshka-\velusqroot: 
the $U \times V$ component covers the kernel polynomial only up to $\ell_l$, 
so all dummy factors necessarily appear in $W$. 
Moreover, $W$ consists solely of even multiples of $P$. 
This enables us to validate each $x$-coordinate of the multiples 
$[2]P, [4]P, \dots, [\tfrac{\ell_r - 1}{2}]P$ 
by checking whether they match the double of their corresponding halves, 
that is, by verifying $xDBL(x([i]P)) = x([2i]P)$.

Depending on the batch size and the velusqrt parameters, in some cases, 
not all odd halves are generated within $U \times V$.
Therefore, the odd points must be computed explicitly in the range
\[
 max\bigl(bs, (\frac{(\ell_r - 1)}{2} - 2 \cdot bs \cdot gs) / 2\bigr),
\]
where $(bs, gs)$ denote the baby-step/giant-step parameters of \velusqroot for 
the $\ell_l$-isogeny.  
This ensures that every even multiple in \( W \) pairs with its half, 
allowing for consistent validation without dummy points. \Cref{alg:matry1414} summarizes the resulting \emph{dummy-free Matryoshka algorithm} adapted to \velusqroot.

\begin{algorithm}
    \caption{Matryoshka 1.414}\label{alg:matry1414}
    \begin{algorithmic}[1]
    \Require A degree $\ell_k$, a batch $[\ell_l, \ldots \ell_r]$, a point $P$ such that $\ell_k \cdot P = \mathcal{O}$ 
                and \velusqroot~parameters ($bs, gs$) for $\ell_l$
    \Ensure The kernel polynomial $h(x)$ for $\phi: E \to E/\langle P \rangle$
    
    \item[]
    \State $b_k \leftarrow \frac{\ell_k - 1}{2}, b_l \leftarrow \frac{\ell_l - 1}{2}, b_r \leftarrow \frac{\ell_r - 1}{2}$
    \State $t \leftarrow b_r - b_l$
    \State Compute multiples according to \velusqroot
    \State Compute odd multiples $[bs+2]P,\dots,[(br - 2*bs*gs)/2]P$ if $bs < (br - 2*bs*gs)/2$
    \State $h(x) \leftarrow 1$
    
    \item[]
    \State Compute \velusqroot~using $(bs,gs)$
    
    \item[]
    \For{$i \in [0, \ldots, b_r - 2*bs*gs]$}
        \State $m \leftarrow \frac{1}{2}x([2*i+2]P)$
        \State $\alpha \leftarrow 1$ if $i \leq b_k - 2*bs*gs$ else $-1$
        \State $\alpha \leftarrow \alpha \cdot$ -\Call{cCompare}{$\xDBL(x([i+1]P))$, $x([2*i+2]P)$} \Comment{-1 if points are equal, else 1.}

        \State $h_1(x) \leftarrow h(x)\cdot(x - m)$
        \State $h_2(x) \leftarrow h(x)\cdot \alpha \cdot m$
        \State $h(x) \leftarrow h_1(x) - h_2(x)$ \Comment{$h$ is multiplied by $x$ when $\alpha = -1$}
    \EndFor
    
    \item[]
    \State \Return $h(x) \leftarrow x^{b_k - b_r} \cdot h(x)$
    \end{algorithmic}
\end{algorithm}

As a result of the $\xDBL$ trick, the overhead of projective Matryoshka 1.414 is just
$2 \cdot \M + xDAC$ for iteration, together with the $((br - 2*bs*gs)/2) - bs$ additionial $\xADD$ to 
compute the missing point halves. 
 
\section{Implementation}
\label{sec:implementation}

We base our implementation on the \dctidh code from \url{https://github.com/PaZeZeVaAt/dCTIDH}, 
which in turn builds on the \texttt{secsidh}\footnote{Publicly available at 
\url{https://github.com/kemtls-secsidh/secsidh}.} implementation~\cite{secsidh}. 
This code incorporates the optimal strategies introduced in~\cite{optstrat} 
to accelerate kernel point computations by balancing the trade-off between 
pushing points through isogenies and computing kernels via DACs. 
In addition, it provides assembly-optimized $\Fp$ arithmetic for the different parameter sets.

We extend this implementation by integrating the new \dacshund parameters for DAC computation 
and by adapting \Cref{alg:matry2} and \Cref{alg:matry1414} to projective space.

\subsection{Batch Configurations}

To determine optimal parameter sets for \textsf{\dctidh}, 
we build on the configurations reported in the original \textsf{\dctidh} work. 
In particular, we focus on the parameter sets \textsf{\dctidh-194} and \textsf{\dctidh-205}, 
which serve as natural starting points and enable direct comparison with their 
non–dummy-free \dctidh counterparts.

Further analysis shows that the small primes $3$, $5$, and $7$ in the set $\{\ell_i\}$ 
severely restrict possible batch structures under \dacshund constraints. 
To address this, we run our greedy search while excluding either $3$, or $3,5,7$ 
from the set $\{\ell_i\}$.  

\Cref{tab:greedyresults} presents the results for the 
\textsf{\dctidh-194} and \textsf{\dctidh-205} parameter sets. 
We evaluate configurations with between $12$ and $20$ batches for each parameter set. 
A complete run over all batch configurations requires approximately 16 hours using 32 threads 
on a server equipped with dual AMD EPYC~7643 processors (\SI{2.3}{\giga\hertz}, 192 threads in total).

\begin{table}[ht]
    \caption{Best greedy results for the \textsf{\dctidh-194} and \textsf{\dctidh-205} parameter sets.}
    \label{tab:greedyresults}
	\centering
	\setlength{\tabcolsep}{1em}
    \begin{tabular}{l c c c c } 
     \toprule
     \emph{variant} & $\ell$ skipped & batches & isogenies & cost  \\ 
     \midrule
     \dctidh-205    &  --       & $15$   & $70$ & $327,942$  \\ 
     \dctidh-194    &  --       & $17$   & $75$ & $334,458$  \\ 
     \midrule
     \dctidh-205 &    $3$     &  $17$  & $73$ & $327,390$  \\ 
     \dctidh-194 &    $3$     &  $14$  & $73$ & $332,920$ \\ 
     \midrule
     \dctidh-205 &  $3,5,7$    & $13$   & $70$ & $334,846$\\ 
     \dctidh-194 &  $3,5,7$    & $13$   & $72$  & $341,526$ \\ 
     \bottomrule
    \end{tabular}
\end{table}

While the performance differences remain within $\approx5\%$, 
our results indicate that the best configuration comes from skipping only the prime~$3$. 
Therefore, we implement dummy-free \dctidh for the parameter sets \textsf{\dctidh-205} and 
\textsf{\dctidh-194} by excluding the $3$-isogeny.  

\begin{remark}
    The greedy search only optimizes the plain cost of isogeny evaluation using optimal strategies.
    Therefore, it does not account for additional, albeit constant, costs in the group action, such as cofactor 
    removal, and a final inversion to return an affine codomain, are not accounted for, explaining the differences.
    to the benchmarks measured in \Cref{tab:implementation}.
\end{remark}

\subsection{Performance}
All benchmarks were performed on an Intel Core i7-6700 (Skylake) processor, 
running Debian~12 with Hyper-Threading and Turbo Boost disabled, and compiled 
using \texttt{gcc}-12.2.0.

\Cref{tab:implementation} compares the results against \dcsidh as only other 
constant-time, dummy-free and deterministic scheme, \ctidh (from the secsidh implementation), 
as well as the relevant \dctidh parameter sets.

\begin{table}[ht]
    \caption{Results of a group action evaluation in multiplications 
    (\M), squarings (\Sq), and additions (\A), and median cycle count (Gcyc) of 10,000 experiments,
    performed on a Skylake CPU.}
    \label{tab:implementation}
	\centering
	\setlength{\tabcolsep}{1em}
    \resizebox{\textwidth}{!}{\begin{tabular}{l  c c c c c } 
     \toprule
     \emph{variant} & \M & \Sq & \A & $\Fp$-mult. & Gcyc \\ \midrule
    \ctidh-2048  &   $287,207\pm21\%$   &  $83,759\pm9\%$  & --   & $370,966\pm17\%$ &  $1.652\pm17\%$ \\
    \dcsidh-2048~\cite{secsidh}  & $1,315,203$     & $227,501$   & --   & $1,542,704$ & $7.039$  \\ 
     \dctidh-2048-205~\cite{dctidh2025} &  $263,545$       & $50,825$    & $465,224$  &  $314,370$ & $1.418$ \\ 
     \dctidh-2048-194~\cite{dctidh2025} &  $266,101$       & $51,258$    & $469,258$  &  $317,359$ & $1.410$ \\ 
     \midrule
     This work (205) &    $303,058$     &  $54,074$  & $560,276$  & $357,132$ & $1.600$ \\ 
     This work (194) &    $307,004$     &  $55,215$  & $553,193$  & $362,219$ & $1.595$ \\ 
     \bottomrule
    \end{tabular}}
\end{table}

\Cref{tab:implementation} compares the cost of the group action across 
different \textsf{CSIDH} implementations. As expected, \dcsidh is by far the most expensive: 
its fully deterministic and dummy-free design results in more than 
$1.5$ million field multiplications and a median cost of 
$7.0$ Gigacycles, making it impractical in comparison with 
other approaches. 

Both parameter sets of \dctidh (194 and 205) are more efficient, 
requiring about $314$--$317$k $\Fp$ multiplications and completing a group action 
in roughly $1.410$--$1.418$ Gigacycles. This confirms that batching and 
WOMBats provide a strong efficiency, albeit at the cost of dummy operations. 

Our dummy-free implementation adds a small overhead compared to \dctidh: 
$358$--$362$k $\Fp$-multiplications and $1.595$--$1.600$ Gigacycles. 
This represents a slowdown of only $12$--$14\%$, while completely eliminating 
dummy multiplications in both DACs and Matryoshka isogenies (when we compare with \dctidh). 
At the same time, we still outperform the original \ctidh by 
about $4\%$, demonstrating the advantages of the \wombat keyspace, even under
the additional \dachshund constraints.

\begin{remark}
    Similar to \dctidh, this work focuses solely on optimizing the group action, 
    which is just one part of a full key exchange. During key generation,
    One also needs to compute a torsion point of order $\prod \ell_i$, and in the
    key derivation step, the order of this point must be validated (which also ensures
    supersingularity). However, excluding the degree 3 speeds up the point search
     and validation by up to 20\% compared to the \dctidh.
    Recent work by Pope, Reijnders, Robert, Sferlazza, and Smith \cite{DBLP:journals/cic/PopeRRSS25}
    used a pairing-based approach for validation, suggesting a possible fourfold speedup. We leave the integration
    of pairing-based validation and point search into the \dctidh-framework as future work.

\end{remark}
 
\section{Conclusion}
\label{sec:conclusion}
We have presented the first \textbf{dummy-free implementation of \dctidh}, 
combining \dacshund with \emph{dummy-free Matryoshka isogenies}. 
Our approach eliminates all dummy operations in both differential addition chains 
and isogeny evaluations, providing the first \dctidh implementation that is 
deterministic, constant-time, and fully dummy-free.
We showed how to adapt the greedy parameter search to incorporate these constraints, 
and identified viable parameter sets for \dctidh-194 and \dctidh-205, 
noting that very small primes such as $3,5,7$ are incompatible with \dacshund.

In our implementation, we report results in \Cref{tab:implementation} using the new 
batching strategy and the Matryoshka~$1.414$ variant. We show that even without dummy isogenies, 
our performance remains close to that of \dctidh. Moreover, we demonstrate an 
improvement of roughly $4\%$ over \ctidh for both our implementations of \dctidh-2048-194 and \dctidh-2048-205.

\ifsubmission

\else
\section*{Acknowledgements}
The authors would like to thank Fabio Campos, 
Krijn Reijnders, and Michael Meyer for their 
discussions and insights on Matryoshka isogenies.
 \fi

\bibliographystyle{plain}
\bibliography{references.bib}

\end{document}